\documentclass[twocolumn,nolinenumbers]{aastex701}

\shorttitle{Two Interstellar Meteor Candidates in the CNEOS Database}
\shortauthors{Cloete \& Loeb}
\newcommand{\vinfsun}{v_{\infty,\odot}}
\newcommand{\vescsun}{v_{\mathrm{esc},\odot}}
\newcommand{\vhel}{v_\odot}
\newcommand{\pbound}{p_\mathrm{bound}}

\newcommand{\kms}{~\mathrm{km\,s^{-1}}}
\newcommand{\okina}{\textquoteleft}
\newcommand{\kmssq}{~\mathrm{km^2\,s^{-2}}}

\graphicspath{{../../../figures/}}

\begin{document}

\title{Two Robust Interstellar Meteor Candidates in the Post-2018 CNEOS Fireball Database}

\author{Richard Cloete}
\affiliation{Astronomy Department, Harvard University, 60 Garden Street, Cambridge, MA, 02138, USA}
\email[]{richardcloete@cfa.harvard.edu}

\author{Abraham Loeb}
\affiliation{Astronomy Department, Harvard University, 60 Garden Street, Cambridge, MA, 02138, USA}
\email[show]{aloeb@cfa.harvard.edu}

\begin{abstract}
We report the identification of two previously unrecognized interstellar meteor candidates in the NASA CNEOS fireball database. Exploiting the empirically calibrated low-discrepancy uncertainty model of Pe\~{n}a-Asensio et al.\ (2025), which characterizes post-2018 CNEOS velocity accuracy at $\sigma_v = 0.55\kms$, $\sigma_\mathrm{RA} = 1.35\degr$, $\sigma_\mathrm{Dec} = 0.84\degr$, we transform CNEOS velocity vectors to heliocentric orbits and assess interstellar candidacy via $10^6$-draw Monte-Carlo simulations.
Two post-2018 events have heliocentric speeds robustly exceeding the escape velocity from the Solar System.
CNEOS-22 (2022-07-28; $6.0\degr$S, $86.9\degr$W; eastern tropical Pacific) has $\vhel = 46.98\kms$, exceeding solar escape by $\langle\Delta\rangle = 5.18 \pm 0.60\kms$ ($z_\Delta = 8.7\sigma$), with heliocentric interstellar speed $\vinfsun = 21.5\kms$.
CNEOS-25 (2025-02-12; $73.4\degr$N, $49.3\degr$E; Barents Sea) has $\vhel = 45.63\kms$, exceeding escape by $\langle\Delta\rangle = 3.22 \pm 0.58\kms$ ($z_\Delta = 5.5\sigma$), with $\vinfsun = 16.9\kms$.
For both events, none of $10^6$ Monte-Carlo realizations yield an orbit that is gravitational bound to the Solar System ($\pbound < 3 \times 10^{-6}$). The adopted error model would need to underestimate the true uncertainties by factors of 5--9 for either candidate's unbound status to be marginal.
\end{abstract}


\section{Introduction}
\label{sec:intro}

The discoveries of 1I/\okina Oumuamua \citep{Meech2017}, 2I/Borisov \citep{Guzik2020}, and 3I/ATLAS \citep{seligman2025discovery} demonstrated that large ($\gtrsim 0.1$~km) interstellar objects transit the inner Solar System.
Although smaller bodies evade telescope detection, they can reveal themselves as fireballs when they enter Earth's atmosphere at speeds exceeding the local escape velocity from the Solar System.

The CNEOS fireball database, maintained by NASA's Jet Propulsion Laboratory, is a global, space-based catalog providing event-level velocity vectors for bolide detections from U.S.\ Government sensors.
These velocity measurements enable computation of heliocentric orbits and testing whether an impactor was gravitationally unbound to the Sun.
However, CNEOS publishes no per-event uncertainties, and the accuracy of reported velocities has varied across sensor generations.

Previous interstellar meteor reports have focused on IM1 from 2014-01-08 \citep{Siraj1} and IM2 from 2017-03-09 \citep{SirajLoeb2022,Pena22}, based on nominal heliocentric speeds exceeding escape velocity.
Independent analysis of IM1 by the U.S. Space Command confirmed its interstellar identification\footnote{\url{https://lweb.cfa.harvard.edu/~loeb/DoD.pdf}}, 
but others raised concerns about the velocity accuracy of pre-2018 CNEOS data \citep{Devillepoix2019}, arguing that without a calibrated uncertainty model, it is impossible to distinguish genuinely unbound trajectories from measurement artifacts.

The situation improved substantially with the empirical calibration of \citet{PenaAsensio2025}, who cross-matched CNEOS events with independent ground-truth networks.
Their analysis reveals two regimes: a high-discrepancy regime for pre-2018 events (speed errors $\sigma_v\sim$6~km\,s$^{-1}$, radiant uncertainties of tens of degrees) and a \textit{low-discrepancy} regime for post-2018 events ($\sigma_v \approx 0.55\kms$, $\sigma_\mathrm{RA} \approx 1.35\degr$, $\sigma_\mathrm{Dec} \approx 0.84\degr$).
This calibration provides, for the first time, an empirically grounded basis for formal statistical assessment of interstellar candidacy.

In this Letter, we report two previously unrecognized post-2018 CNEOS fireballs whose heliocentric speeds robustly exceed the solar escape velocity under the calibrated low-discrepancy error model.
Both events remain unbound to the Solar System in $10^6$ Monte-Carlo realizations, with many-sigma margins above the escape speed.

\section{Data and Analysis}
\label{sec:methods}

\subsection{CNEOS Data and Ephemerides}

We analyze the complete CNEOS fireball catalog, retaining events with reported latitude, longitude, altitude, and three-component velocity vectors.
For each event, the Earth-fixed (ITRS) velocity is transformed to inertial geocentric (GCRS) coordinates using Astropy \citep{Astropy2022}, accounting for Earth rotation, precession, nutation, and polar motion.
Earth's gravitational influence is removed via a two-body hyperbolic model to obtain the geocentric excess velocity $\mathbf{v}_{\infty,\oplus}$.
The heliocentric velocity is then $\mathbf{v}_\odot = \mathbf{v}_{\infty,\oplus} + \mathbf{v}_E$, where $\mathbf{v}_E$ is the Earth's heliocentric velocity from JPL Horizons \citep{Giorgini1996} at the event time.
An event is classified as an interstellar candidate when $|\mathbf{v}_\odot| > \vescsun(r_E) = \sqrt{2GM_\odot / r_E} \approx 42\kms$, where $r_E$ is the Earth-Sun separation at the event time.

\subsection{Uncertainty Propagation}

We adopt the low-discrepancy uncertainties from \citet{PenaAsensio2025}: $\sigma_v = 0.55\kms$ (speed), $\sigma_\mathrm{RA} = 1.35\degr$, $\sigma_\mathrm{Dec} = 0.84\degr$ (radiant direction), and restrict our candidacy claims to the post-2018 era where these values apply.

For each candidate, we perform $N = 10^6$ Monte-Carlo realizations.
In each draw, the geocentric speed magnitude is perturbed by $\mathcal{N}(0, \sigma_v)$ (normal distribution) and the velocity direction is perturbed on the tangent plane with per-axis angular uncertainty $\sigma_{\theta,\mathrm{axis}} = \sqrt{(\sigma_\mathrm{RA}^2 + \sigma_\mathrm{Dec}^2)/2} \approx 1.12\degr$.
The full orbit computation chain is re-evaluated for each realization.

We report two statistics: (1)~the fraction of realizations yielding unbound orbits, with the rule-of-three providing a 95\% upper limit on $\pbound$ when zero bound draws are observed; and (2)~the continuous margin $\Delta = (\vhel - \vescsun)$, whose significance $z_\Delta = \langle\Delta\rangle / \sigma_\Delta$ expresses the mean excess above escape in units of its Monte-Carlo scatter.

\section{Results}
\label{sec:results}

\begin{figure}
\centering
\includegraphics[width=\columnwidth]{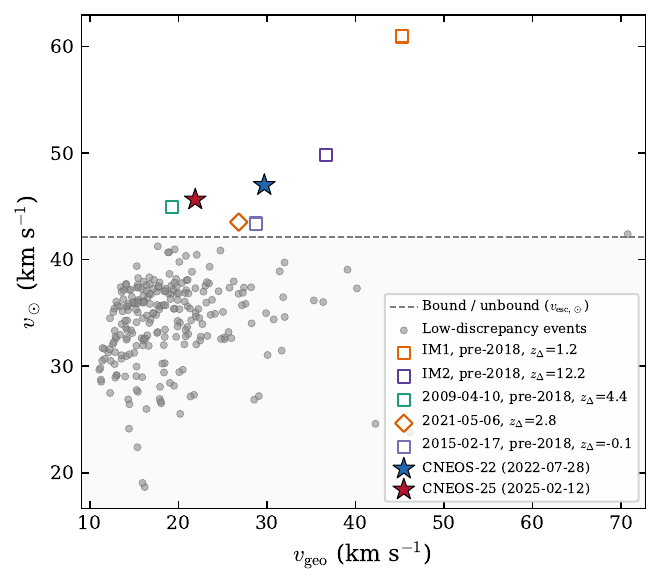}
\caption{Heliocentric speed $\vhel$ versus geocentric speed $v_\mathrm{geo}$ for CNEOS fireballs with complete velocity vectors. The horizontal dashed line marks the bound/unbound boundary at the solar escape speed $\vescsun \approx 42\kms$. Grey circles are low-discrepancy events with bound nominal orbits. Several events lie above the boundary at their nominal velocities (individual legend entries): pre-2018 events (open squares), including IM1 and IM2, and a post-2018 marginal event (open diamond). CNEOS-22 (2022-07-28) and CNEOS-25 (2025-02-12) (colored stars) are the only post-2018 events that remain robustly unbound across all $10^6$ Monte-Carlo realizations ($z_\Delta > 5$; $\pbound < 3 \times 10^{-6}$). 
}
\label{fig:scatter}
\end{figure}

Restricting attention to the post-2018 low-discrepancy era, two events emerge as robust interstellar candidates.
Both have positive heliocentric specific energy at their nominal velocities and remain unbound to the Solar System in all $10^6$ Monte-Carlo realizations.
Neither has been previously identified as an interstellar candidate.
Figure~\ref{fig:scatter} places both events in context of the full post-2018 catalog.

\subsection{CNEOS-22: 2022-07-28}

This fireball occurred at 01:36:07 UTC over the eastern tropical Pacific Ocean ($6.0\degr$S, $86.9\degr$W), approximately 600~km west of Peru, at an altitude of 37.5~km.
Its impact energy of 0.69~kt places it comfortably above the 0.45~kt secondary low-discrepancy threshold.
The CNEOS velocity components are $(v_x, v_y, v_z) = (-17.1, +23.5, -7.2)\kms$.

After coordinate transformation, Earth gravity removal, and heliocentric vector addition, the heliocentric speed is $\vhel = 46.98\kms$, exceeding the solar escape speed of $\vescsun = 41.79\kms$ by 5.19~km\,s$^{-1}$.
The heliocentric excess speed is $\vinfsun = 21.5\kms$ and the specific orbital energy is $\varepsilon_\odot = +230\kmssq$.

Under the Monte-Carlo analysis, $\langle\vhel\rangle = 46.98 \pm 0.60\kms$ and the margin $\langle\Delta\rangle = 5.18 \pm 0.60\kms$ ($z_\Delta = 8.7$).
None of $10^6$ draws yield an orbit bound to the Solar System.

The impact energy and speed imply a bolide with a 
mass of $6.4\times 10^6~{\rm g}$ and a radius of $91~{\rm cm}\times (\rho/2~{\rm g~cm^{-3}})^{-1/3}$ for a solid density $\rho$. The air's ram-pressure at the fireball's peak-brightness altitude was 5.2 MPa.

\subsection{CNEOS-25: 2025-02-12}

This fireball occurred at 04:33:39 UTC over the Barents Sea ($73.4\degr$N, $49.3\degr$E), between Novaya Zemlya and Franz Josef Land in the high Arctic, at 42.0~km altitude.
Its impact energy is 0.13~kt; it qualifies for the low-discrepancy regime by its post-2018 date.
The CNEOS velocity components are $(v_x, v_y, v_z) = (+9.5, +19.6, -1.9)\kms$.

The heliocentric speed is $\vhel = 45.63\kms$, exceeding the escape speed of $\vescsun = 42.40\kms$ by 3.23~km\,s$^{-1}$.
The heliocentric excess speed is $\vinfsun = 16.9\kms$ and $\varepsilon_\odot = +142\kmssq$.

Under the Monte-Carlo analysis, $\langle\vhel\rangle = 45.62 \pm 0.58\kms$ and $\langle\Delta\rangle = 3.22 \pm 0.58\kms$ ($z_\Delta = 5.5$).
None of $10^6$ draws yield a bound orbit.

The impact energy and speed imply a bolide with a 
mass of $2.2\times 10^6~{\rm g}$ and a radius of $64~{\rm cm}\times (\rho/2~{\rm g~cm^{-3}})^{-1/3}$ for a solid density $\rho$. The air's ram-pressure at the fireball's peak-brightness altitude was 1.5 MPa.

\subsection{Summary}

\begin{figure*}
\centering
\includegraphics[width=\textwidth]{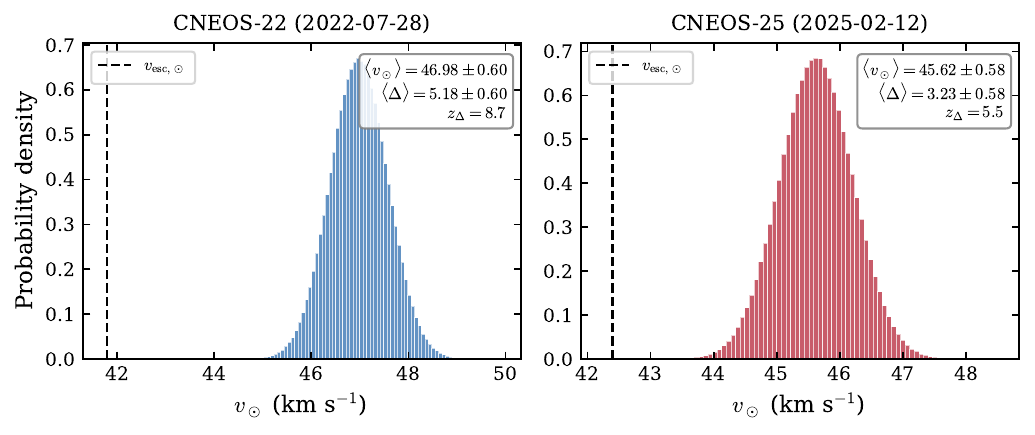}
\caption{Monte-Carlo distributions of heliocentric speed $\vhel$ for CNEOS-22 (2022-07-28; \textit{left}) and CNEOS-25 (2025-02-12; \textit{right}), based on $10^6$ realizations of the \citet{PenaAsensio2025} low-discrepancy uncertainty model. The dashed vertical line marks the solar escape speed $\vescsun$ at each event's heliocentric distance. In both cases the entire distribution lies above Solar System escape, with no bound realizations observed. Inserts provide the mean heliocentric speed (in km~s$^{-1}$), the mean margin above escape $\langle\Delta\rangle$ (in km~s$^{-1}$), and its statistical significance in units of standard deviations, $z_\Delta$.}
\label{fig:mc_hist}
\end{figure*}

Table~\ref{tab:summary} presents the key properties of both candidates.
For each, the 95\% upper limit on the bound probability is $\pbound < 3 \times 10^{-6}$ (rule of three for 0 in $10^6$ trials).
The margin significances of $8.7\sigma$ and $5.5\sigma$ quantify how deeply into the unbound regime each event lies, independent of the Monte-Carlo sample size.
The Monte-Carlo velocity distributions are shown in Figure~\ref{fig:mc_hist}.

\begin{deluxetable*}{lcc}
\tablecaption{Properties of the two interstellar meteor candidates.\label{tab:summary}}
\tablehead{
\colhead{Property} & \colhead{CNEOS-22 (2022-07-28)} & \colhead{CNEOS-25 (2025-02-12)}
}
\startdata
Date/time (UTC) & 2022-07-28 01:36:07 & 2025-02-12 04:33:39 \\
Latitude & $-6.0\degr$ (S) & $+73.4\degr$ (N) \\
Longitude & $-86.9\degr$ (W) & $+49.3\degr$ (E) \\
Altitude (km) & 37.5 & 42.0 \\
Impact energy (kt) & 0.69 & 0.13 \\
Location & Pacific Ocean & Arctic (Barents Sea) \\
$\vhel$ (km\,s$^{-1}$) & 46.98 & 45.63 \\
$\vescsun$ (km\,s$^{-1}$) & 41.79 & 42.40 \\
$\vinfsun$ (km\,s$^{-1}$) & 21.5 & 16.9 \\
$\varepsilon_\odot$ (km$^2$\,s$^{-2}$) & $+230$ & $+142$ \\
$k_\mathrm{bound}$ / $N$ & 0 / $10^6$ & 0 / $10^6$ \\
$\pbound$ (95\% UL) & $< 3\times10^{-6}$ & $< 3\times10^{-6}$ \\
$\langle\Delta\rangle \pm \sigma_\Delta$ (km\,s$^{-1}$) & $5.18 \pm 0.60$ & $3.22 \pm 0.58$ \\
$z_\Delta$ & 8.7 & 5.5 \\
\enddata
\end{deluxetable*}

\section{Discussion}
\label{sec:discussion}

\subsection{Robustness and Systematics}

\begin{figure}
\centering
\includegraphics[width=\columnwidth]{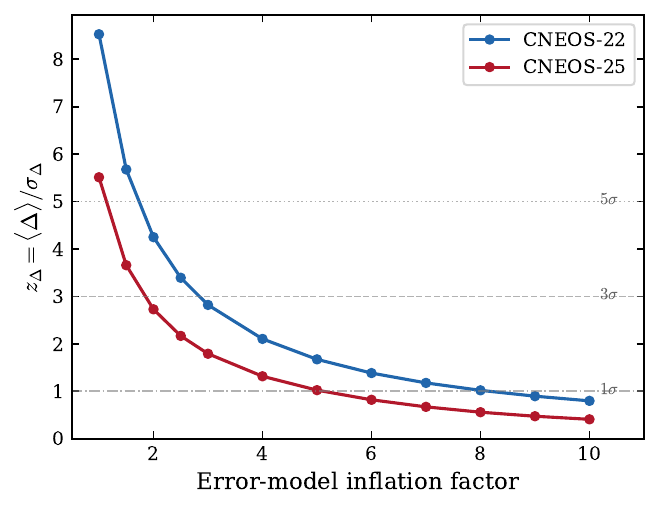}
\caption{Sensitivity of the interstellar classification to systematic underestimation of CNEOS velocity uncertainties. The margin significance $z_\Delta$ is plotted as a function of a multiplicative inflation factor applied uniformly to all three uncertainty components ($\sigma_v$, $\sigma_\mathrm{RA}$, $\sigma_\mathrm{Dec}$). CNEOS-22 (2022-07-28) remains above $3\sigma$ until the errors are inflated by $\sim$3$\times$, and above $1\sigma$ until $\sim$9$\times$. CNEOS-25 (2025-02-12) crosses $3\sigma$ near $\sim$2$\times$ and $1\sigma$ near $\sim$5.5$\times$.}
\label{fig:robustness}
\end{figure}

Our results are conditional on the \citet{PenaAsensio2025} low-discrepancy error model.
If the true uncertainties for these events are larger than adopted, the inferred bound fractions would increase.
However, the error model would need to underestimate the true uncertainties by a factor of $\sim$8.7 for CNEOS-22 and $\sim$5.5 for CNEOS-25 before the mean margin drop to $1\sigma$---inflation factors inconsistent with the calibration data.
Figure~\ref{fig:robustness} quantifies this: both candidates maintain $z_\Delta > 3$ until the uncertainties are inflated by factors of $\sim$2--3, and CNEOS-22 remains above $1\sigma$ even at $\sim$9$\times$ inflation.

The CNEOS velocities are measured at peak brightness after some atmospheric deceleration.
The reported speeds are therefore \textit{lower bounds} on the true entry speeds, and a correction would only strengthen the unbound classifications.
Similarly, the two-body approximations used for Earth gravity removal and the solar binding test introduce sub-percent velocity corrections, negligible compared to the km\,s$^{-1}$ measurement uncertainties.

\subsection{Comparison with Known Interstellar Objects}

The heliocentric interstellar speeds of the two candidates---$\vinfsun = 21.5\kms$ and $16.9\kms$---sample the low end of the velocity distribution expected for objects from the Galactic solar neighborhood.
For comparison, 1I/\okina Oumuamua entered the Solar System with $\vinfsun \approx 26\kms$ \citep{Meech2017} and 2I/Borisov with $\vinfsun \approx 32\kms$ \citep{Guzik2020}, whereas 3I/ATLAS entered with $\vinfsun \approx 58\kms^-1$ \citep{seligman2025discovery}.
The lower excess speeds of the fireball candidates are qualitatively consistent with the expectation that smaller, fainter impactors are preferentially detected at lower encounter speeds, as long as their atmospheric luminosity is sufficient to trigger space-based sensors.

\subsection{Material Recovery Prospects}

CNEOS-22 occurred over the open Pacific at $\sim$3--4~km depth, presenting deep-ocean recovery challenges similar to the IM1 expedition \citep{Loeb2024}.
Its higher impact energy (0.69~kt) suggests a more massive impactor with potentially more recoverable debris than CNEOS-25. However,
CNEOS-25's Arctic location (Barents Sea) offers the advantage of a shallow continental shelf ($\sim$200--400~m) but the constraints of sea-ice logistics.
Crucially, CNEOS-25 is recent (February 2025), and rapid mobilization could maximize recovery prospects before material redistribution by ice drift and currents.

\subsection{Implications for Interstellar Meteoroid Flux}

Two detections over $\sim$7~years of post-2018 monitoring implies an Earth-impact rate on the order of $\sim$0.3~events~yr$^{-1}$ for meter-scale interstellar bolides at the CNEOS detection threshold, consistent with past expectations~\citep{SirajLoeb2022,Pena2,jewitt2023interstellar}.

\section{Summary}
\label{sec:summary}

We report two previously unrecognized interstellar meteor candidates in the post-2018 CNEOS fireball database, identified by computing heliocentric orbits from space-based velocity measurements and propagating measurement uncertainty through Monte-Carlo simulations calibrated to the empirical low-discrepancy error model of \citet{PenaAsensio2025}:

\begin{enumerate}
\item \textbf{CNEOS-22} (2022-07-28; eastern tropical Pacific): $\vhel = 46.98\kms$, $\vinfsun = 21.5\kms$, $z_\Delta = 8.7\sigma$ above escape.
\item \textbf{CNEOS-25} (2025-02-12; Barents Sea, Arctic): $\vhel = 45.63\kms$, $\vinfsun = 16.9\kms$, $z_\Delta = 5.5\sigma$ above escape.
\end{enumerate}

For both events, none of $10^6$ Monte-Carlo realizations yield a bound heliocentric orbit ($\pbound < 3 \times 10^{-6}$ at 95\% confidence), and the adopted error model would need to underestimate the true uncertainties by factors of 5--9 to render either candidate marginal.
These are the strongest interstellar meteor candidates yet identified in the calibrated era of the CNEOS record.
Follow-up N-body orbital reconstruction, fall-location modeling, and---where feasible---material recovery expeditions are warranted, particularly for the recent CNEOS-25 event.

\begin{acknowledgments}
This work was supported by the Galileo Project at Harvard University.
The reported calculations made use of Astropy \citep{Astropy2022}, the JPL Horizons system \citep{Giorgini1996}, NumPy, and SciPy.
\end{acknowledgments}

\software{
Astropy \citep{Astropy2022},
NumPy \citep{numpy2020},
SciPy \citep{scipy2020},
Astroquery \citep{Ginsburg2019}
}

\bibliography{references}{}

@article{Meech2017,
  title={A brief visit from a red and extremely elongated interstellar asteroid},
  author={Meech, Karen J and Weryk, Robert and Micheli, Marco and Kleyna, Jan T and Hainaut, Olivier R and Jedicke, Robert and Wainscoat, Richard J and Chambers, Kenneth C and Keane, Jacqueline V and Petric, Andreea and others},
  journal={Nature},
  volume={552},
  number={7685},
  pages={378--381},
  year={2017},
  publisher={Nature Publishing Group UK London}
}

@article{Guzik2020,
  title={Initial characterization of interstellar comet 2I/Borisov},
  author={Guzik, Piotr and Drahus, Micha{\l} and Rusek, Krzysztof and Waniak, Wac{\l}aw and Cannizzaro, Giacomo and Pastor-Marazuela, In{\'e}s},
  journal={Nature Astronomy},
  volume={4},
  number={1},
  pages={53--57},
  year={2020},
  publisher={Nature Publishing Group UK London}
}

@article{SirajLoeb2022,
  title={Interstellar meteors are outliers in material strength},
  author={Siraj, Amir and Loeb, Abraham},
  journal={The Astrophysical Journal Letters},
  volume={941},
  number={2},
  pages={L28},
  year={2022b},
  publisher={IOP Publishing}
}

@article{Devillepoix2019,
  title={Observation of metre-scale impactors by the Desert Fireball Network},
  author={Devillepoix, Hadrien AR and Bland, Philip A and Sansom, Eleanor K and Towner, Martin C and Cup{\'a}k, Martin and Howie, Robert M and Hartig, Benjamin AD and Jansen-Sturgeon, Trent and Cox, Morgan A},
  journal={Monthly Notices of the Royal Astronomical Society},
  volume={483},
  number={4},
  pages={5166--5178},
  year={2019},
  publisher={Oxford University Press}
}

@article{PenaAsensio2025,
  title={Error dependencies in the space-based CNEOS fireball database},
  author={Pe{\~n}a-Asensio, Eloy and Socas-Navarro, Hector and Seligman, Darryl Z},
  journal={Astronomy \& Astrophysics},
  volume={701},
  pages={A202},
  year={2025a},
  publisher={EDP Sciences}
}

@ARTICLE{Loeb2024,
       author = {{Loeb}, A. and {Jacobsen}, S.~B. and {Tagle}, R. and {Adamson}, T. and {Bergstrom}, S. and {Cloete}, R. and {Cohen}, S. and {Domine}, Laura and {Fu}, H. and {Hoskinson}, C. and {Hyung}, E. and {Kelly}, M. and {Lard}, E. and {Laukien}, F. and {Lem}, J. and {McCallum}, R. and {Millsap}, R. and {Parendo}, C. and {Pataev}, M.~I. and {Peddeti}, C. and {Pugh}, J. and {Samuha}, S. and {Sasselov}, D.~D. and {Schlereth}, M. and {Siler}, J. and {Siraj}, A. and {Smith}, P.~M. and {Taylor}, J. and {Weed}, R. and {Wright}, A. and {Wynn}, J.},
        title = "{Chemical Classification of Spherules Recovered From The Pacific Ocean Site of The CNEOS 2014-01-08 (IM1) Bolide}",
      journal = {Chemical Geology},
         year = 2024,
        month = dec,
       volume = {670},
          eid = {122415},
        pages = {122415},
          doi = {10.1016/j.chemgeo.2024.122415},
archivePrefix = {arXiv},
       eprint = {2401.09882},
 primaryClass = {astro-ph.EP},
       adsurl = {https://ui.adsabs.harvard.edu/abs/2024ChGeo.67022415L}
}

@inproceedings{Giorgini1996,
  title={JPL's on-line solar system data service},
  author={Giorgini, JD and Yeomans, DK and Chamberlin, AB and Chodas, PW and Jacobson, RA and Keesey, MS and Lieske, JH and Ostro, SJ and Standish, EM and Wimberly, RN},
  booktitle={AAS/Division for Planetary Sciences Meeting Abstracts\# 28},
  volume={28},
  pages={25--04},
  year={1996}
}

@article{Astropy2022,
  title={The Astropy Project: sustaining and growing a community-oriented open-source project and the latest major release (v5. 0) of the core package},
  author={Astropy Collaboration and Price-Whelan, Adrian M and Lim, Pey Lian and Earl, Nicholas and Starkman, Nathaniel and Bradley, Larry and Shupe, David L and Patil, Aarya A and Corrales, Lia and Brasseur, CE and others},
  journal={The Astrophysical Journal},
  volume={935},
  number={2},
  pages={167},
  year={2022},
  publisher={The American Astronomical Society}
}

@article{Ginsburg2019,
  title={Astroquery: an astronomical web-querying package in Python},
  author={Ginsburg, Adam and Sip{\H{o}}cz, Brigitta M and Brasseur, CE and Cowperthwaite, Philip S and Craig, Matthew W and Deil, Christoph and Groener, Austen M and Guillochon, James and Guzman, Giannina and Liedtke, Simon and others},
  journal={The Astronomical Journal},
  volume={157},
  number={3},
  pages={98},
  year={2019},
  publisher={The American Astronomical Society}
}

@article{numpy2020,
  title={Array programming with NumPy},
  author={Harris, Charles R and Millman, K Jarrod and Van Der Walt, St{\'e}fan J and Gommers, Ralf and Virtanen, Pauli and Cournapeau, David and Wieser, Eric and Taylor, Julian and Berg, Sebastian and Smith, Nathaniel J and others},
  journal={nature},
  volume={585},
  number={7825},
  pages={357--362},
  year={2020},
  publisher={Nature Publishing Group UK London}
}

@article{scipy2020,
  title={SciPy 1.0: fundamental algorithms for scientific computing in Python},
  author={Virtanen, Pauli and Gommers, Ralf and Oliphant, Travis E and Haberland, Matt and Reddy, Tyler and Cournapeau, David and Burovski, Evgeni and Peterson, Pearu and Weckesser, Warren and Bright, Jonathan and others},
  journal={Nature methods},
  volume={17},
  number={3},
  pages={261--272},
  year={2020},
  publisher={Nature Publishing Group US New York}
}

@article{seligman2025discovery,
  title={Discovery and preliminary characterization of a third interstellar object: 3I/ATLAS},
  author={Seligman, Darryl Z and Micheli, Marco and Farnocchia, Davide and Denneau, Larry and Noonan, John W and Hsieh, Henry H and Santana-Ros, Toni and Tonry, John and Auchettl, Katie and Conversi, Luca and others},
  journal={The Astrophysical Journal Letters},
  volume={989},
  number={2},
  pages={L36},
  year={2025},
  publisher={The American Astronomical Society}
}

@article{jewitt2023interstellar,
  author  = {Jewitt, David and Seligman, Darryl Z.},
  title   = {The Interstellar Interlopers},
  journal = {Annual Review of Astronomy and Astrophysics},
  year    = {2023},
  volume  = {61},
  pages   = {197--236},
  doi     = {10.1146/annurev-astro-071221-054221}
}

@ARTICLE{Siraj1,
       author = {{Siraj}, Amir and {Loeb}, Abraham},
        title = "{A Meteor of Apparent Interstellar Origin in the CNEOS Fireball Catalog}",
      journal = {\apj},
         year = {2022a},
        month = nov,
       volume = {939},
       number = {1},
          eid = {53},
        pages = {53},
          doi = {10.3847/1538-4357/ac8eac},
       adsurl = {https://ui.adsabs.harvard.edu/abs/2022ApJ...939...53S}
}

@ARTICLE{Pena2,
       author = {{Pe{\~n}a-Asensio}, E. and {Seligman}, D.~Z.},
        title = "{The interstellar flux gap: From dust to kilometer-scale objects}",
      journal = {\aap},
         year = {2025b},
        month = dec,
       volume = {704},
          eid = {L1},
        pages = {L1},
          doi = {10.1051/0004-6361/202557337},
archivePrefix = {arXiv},
       eprint = {2511.01957},
 primaryClass = {astro-ph.EP},
       adsurl = {https://ui.adsabs.harvard.edu/abs/2025A&A...704L...1P}
}

@ARTICLE{Pena22,
       author = {{Pe{\~n}a-Asensio}, Eloy and {Trigo-Rodr{\'\i}guez}, Josep M. and {Rimola}, Albert},
        title = "{Orbital Characterization of Superbolides Observed from Space: Dynamical Association with Near-Earth Objects, Meteoroid Streams, and Identification of Hyperbolic Meteoroids}",
      journal = {\aj},
         year = 2022,
        month = sep,
       volume = {164},
       number = {3},
          eid = {76},
        pages = {76},
          doi = {10.3847/1538-3881/ac75d2},
archivePrefix = {arXiv},
       eprint = {2206.03115},
 primaryClass = {astro-ph.EP},
       adsurl = {https://ui.adsabs.harvard.edu/abs/2022AJ....164...76P}
}
\bibliographystyle{aasjournalv7}

\end{document}